\documentclass{appolb}
\usepackage{graphicx}
\usepackage{xcolor}
\usepackage{cite}

\begin{document}
\title{ Production yield and azimuthal anisotropy measurements of strange hadrons from BES at STAR %
\thanks{Quark Matter-2022}%
}
\author{Aswini Kumar Sahoo (For the STAR Collaboration)
\address{Indian Institute of Science Education and Research (IISER), Berhampur, India \\
Email: aswinikumar.aks96@gmail.com/aswini96@rcf.rhic.bnl.gov }
}

\maketitle
\begin{abstract}
\textbf{Abstract:} We report the production and azimuthal anisotropy measurements of strange and multi-strange hadrons at STAR BES energies. The $\Lambda/K^0_s$ ratio is reported at 3 GeV and observed to increase faster with transverse momentum than that at higher energies. The number-of-constituent quark (NCQ) scaling of $v_2$ has been studied at 19.6 GeV (BES-II). The NCQ scaling holds for particles and anti-particles, which can be considered as an evidence of partonic collectivity. The production of $K^{*0}$ resonance is also reported for 7.7-39 GeV (BES-I) and the $K^{*0}/K$ ratio suggests that hadronic re-scattering dominates over regeneration in central A+A collisions. Using the $K^{*0}/K$ ratio, we also report the lower limit of hadronic phase lifetime $( t_{kin} - t_{chem} )$.
\end{abstract}
  
\section{Introduction}

Searching for the onset of the deconfinement is one of the main motivations of the Beam Energy Scan (BES) program at RHIC. The production yield and azimuthal anisotropy of (multi-)strange hadrons is considered a good probe to study the properties of the matter produced in heavy-ion collisions.
The (multi-)strange hadrons are expected to freeze out earlier than other light hadrons, such as $\pi, K, p$ \cite{vanHecke:1998yu}. They are also expected to have smaller hadronic interaction cross sections compared to non-strange hadrons~\cite{PhysRevLett.54.1122}. Hence the production of (multi-)strange hadrons should not be strongly affected by the later stage of heavy-ion collisions.



 Resonances, like $K^{*0}(892)$, are short-lived particles produced in high energy collisions. $K^{*0}$, having a smaller lifetime ($\sim$ 4.16 fm/$c$) than the medium ($\sim$ 10 fm/$c$), is expected to be sensitive to the dynamics in the hadronic phase. In between chemical (CFO) and kinetic (KFO) freeze out, the daughter particles of $K^{*0}$(892) could undergo re-scatting and regeneration. The final yield of the $K^{*0}$(892) depends on the interplay of these effects, and can be used to study the hadronic phase of heavy-ion collisions ~\cite{STAR:2002npn}. 

\section{Data Sets and Analysis Details}
 
 In these proceedings, we report strange hadrons yield and elliptic flow in Au+Au collisions at $\sqrt{s_{NN}}$ = 3 and 19.6 GeV, accumulated by the STAR experiment in 2018 and 2019 as part of the RHIC BES-II program.
The $K^{*0}$ spectrum analysis is performed using Au+Au collisions at $\sqrt{s_{NN}}$ = 7.7-39 GeV collected in 2010, 2011 and 2014. 
For particle identification both the Time Projection Chamber (TPC) and the Time Of Flight (TOF) detector are used. In BES-II, the TPC detector has been upgraded for better momentum resolution and wider pseudo-rapidity coverage ($|\eta|< 1.5$). 




\section{Results}

\subsection{Probing partonic phase with strange hadrons }

\subsubsection{$\Lambda/K^0_s$ ratio}

\begin{figure*}[!ht]
\begin{center}
\includegraphics[height = 0.7\textwidth]{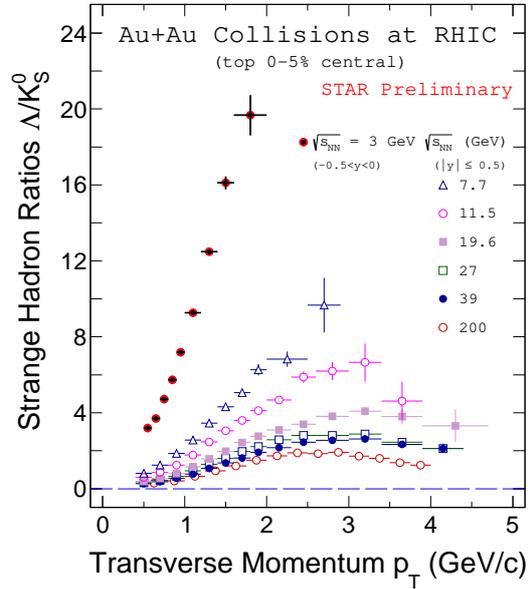}
\caption{ $\Lambda/K^0_s$ as a function of $p_T$ in Au+Au collisions at various beam energies~\cite{PhysRevLett.108.072301}~\cite{PhysRevC.102.034909}. The bars indicate statistical uncertainties only . }
\label{ fig-ks0-lambda-ratio}
\end{center}
\end{figure*}

The baryon-to-meson ratio can be used to investigate the particle production mechanism in heavy-ion collisions. Figure~\ref{ fig-ks0-lambda-ratio} represents the variation of $\Lambda/K^0_s$ as a function of transverse momentum ($p_T$) in central Au+Au collision at various beam energies. We observe that $\Lambda/K^0_s$ increases faster with $p_T$ at 3 GeV compared to higher energies. According to the thermal model:

\begin{equation}
\frac{N(\Lambda)}{N(K^0_s)} \propto exp(\frac{(1-\sigma_s)}{T}) \hspace{1cm} where \hspace{0.3cm}  \sigma_s = \frac{\mu_s}{\mu_B} \hspace{0.5cm}.
 \label{eqn-thermal-ratio}
\end{equation} 
Here T is the temperature and  $\mu_s$ and $\mu_B$ are the strangeness chemical potential and baryon chemical potential respectively. Since there is a significant difference in chemical potential at lower energies compared to higher energies, this trend could be chemical potential driven.


\subsubsection{Elliptic flow $(v_2)$}

\begin{figure*}[!ht]
\begin{center}
\includegraphics[scale=0.7]{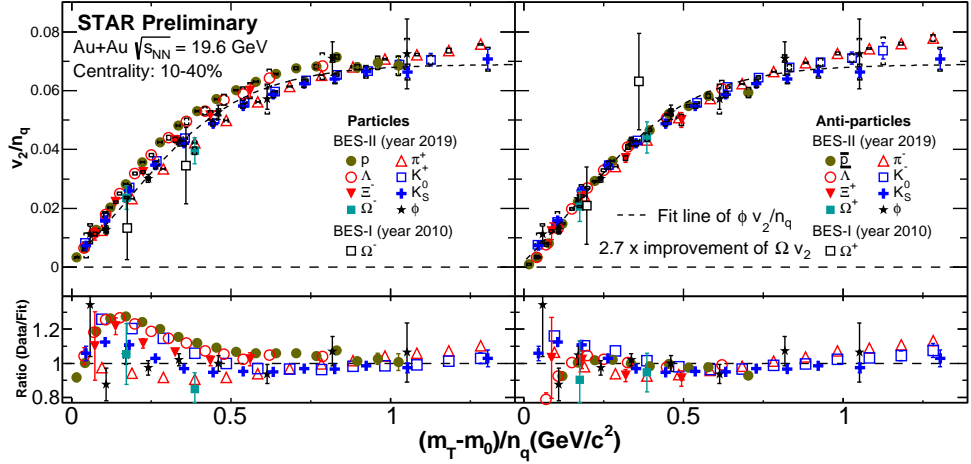}
\caption{The elliptic flow $(v_2)$ scaled by the number of constituent quarks ($n_q$) as a
function of $(m_T-m_0)/n_q$ for particles and their corresponding anti-particles in Au+Au collisions at 19.6 GeV (BES-II) for 10-40\% centrality. The bars and caps indicate statistical and systematic uncertainties respectively.}
\label{ fig-elliptic-flow }
\end{center}
\end{figure*}

In the overlap region of two colliding nuclei the pressure gradient is different in different direction that leads to momentum space anisotropy. This anisotropy is the main cause for the development of elliptic flow ($v_2$). Hence $v_2$ is sensitive to the initial dynamics of heavy-ion collisions. Figure~\ref{ fig-elliptic-flow } shows $v_2$ divided by the number of constituent quarks $(n_q)$ as a function of $(m_T-m_0)/n_q$, where $m_T= \sqrt{(p^2_T + m^2_0)}$ is the transverse mass and $m_0$ is the rest mass of the hadron, at $\sqrt{s_{NN}}$ = 19.6 GeV (BES-II) for 10-40\% centrality. The NCQ scaling holds within 20\% for particles and anti-particles, which could be considered as a signature of partonic collectivity~\cite{PhysRevLett.91.092301}. The scaling holds better for anti-particles than for the particles, which might be due to the transported quark effect.

\subsection{Probing hadronic phase with $K^{*0}$ resonance}

\subsubsection{$K^{*0}/K$ ratio and hadronic phase lifetime}

The decay daughters of $K^{*0}$ (i.e $\pi$ and K) may re-scatter with other particles during the hadronic phase of heavy-ion collisions. Meanwhile, pions and kaons may regenerate $K^{*0}$ via pseudo-elastic scattering. So the $K^{*0}/K$ ratio can be used to probe the relative contributions of these effects. 

\begin{figure*}[!ht]
\begin{center}
\includegraphics[scale=0.31]{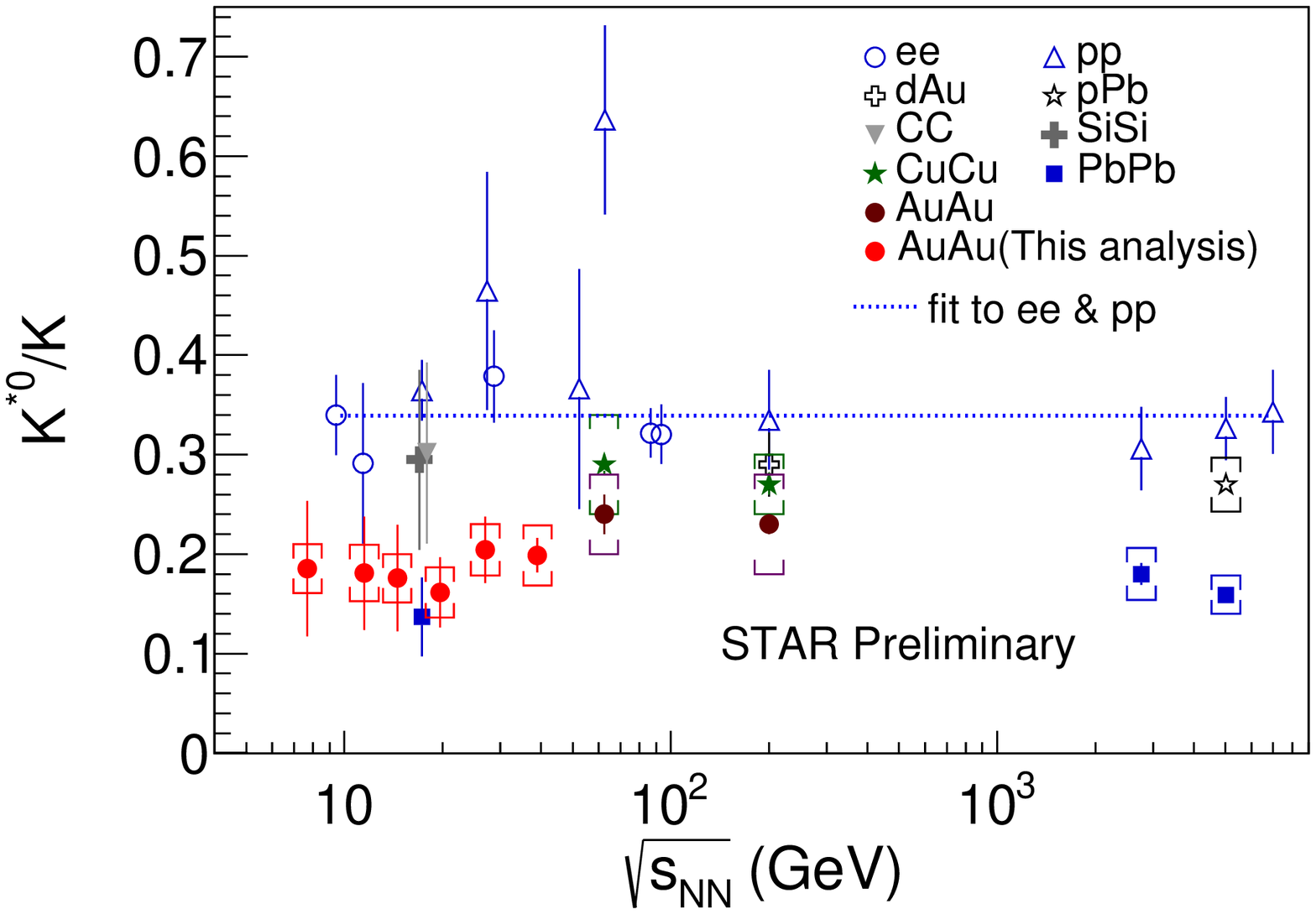}
\includegraphics[scale=0.31]{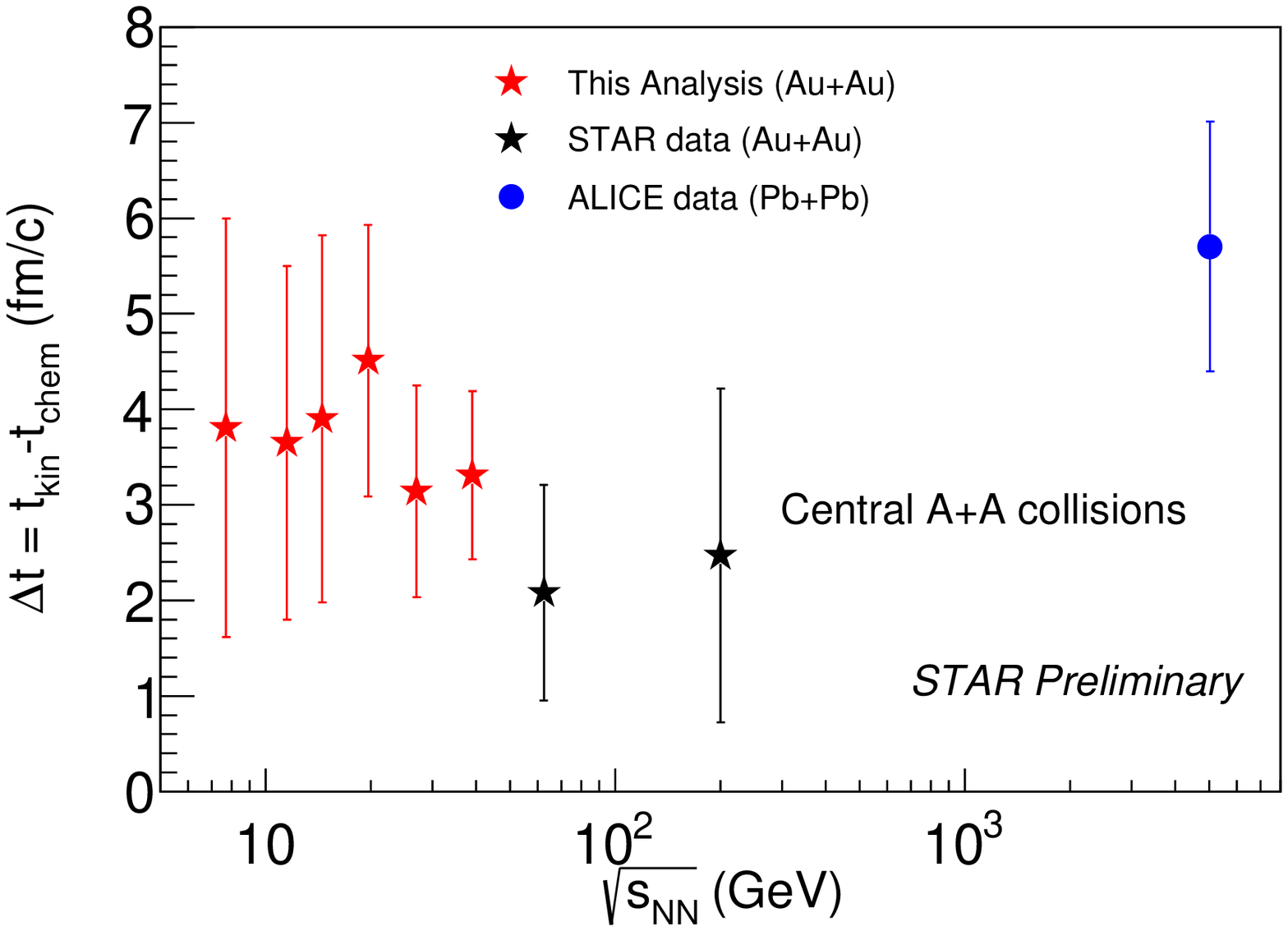}
\caption{Left panel: $K^{*0}/K$ as a function of the collision energy~\cite{ARGUS:1993ggm,Pei:1996kq,Hofmann:1988gy,SLD:1998coh,Aguilar-Benitez:1991hzq,STAR:2004bgh, AnnecyLAPP-CERN-CollegedeFrance-Dortmund-Heidelberg-Warsaw:1981whv,AxialFieldSpectrometer:1982btk,PHENIX:2014kia,STAR:2010avo,STAR:2008twt,NA49:2011bfu,ALICE:2016sak,ALICE:2021uyz,ALICE:2014jbq,ALICE:2017ban,ALICE:2019xyr}. Here the $K^{*0}/K$ represents $(K^{*0}+\overline{K^{*0}})/(K^++K^-)$. The bars and caps indicate statistical and systematic uncertainties respectively. Right panel: Hadronic phase lifetime $(\Delta t)$ as a function of the collision energy. The result is compared with previous STAR~\cite{STAR:2004bgh,STAR:2010avo} and ALICE~\cite{ALICE:2014jbq,ALICE:2017ban,ALICE:2019xyr} results. The error bars are the quadratic sum of the statistical and systematic uncertainties. }
\label{ fig-kstarbyk-ratio}
\end{center}
\end{figure*}

In the left panel of Fig.~\ref{ fig-kstarbyk-ratio} we have shown $K^{*0}/K$ as a function of the collision energy. Here we can see the ratio in central A+A collisions is smaller than in elementary (e+e or p+p)  collisions, indicating that re-scattering might be dominant over re-generation in central A+A collisions.


We can also use the $K^{*0}/K$ ratio to extract the lower limit of hadronic phase lifetime~\cite{STAR:2002npn} following~\cite{Singha:2015fia}, i.e 
\begin{equation}
\bigg(\frac{K^{*0}}{K}\bigg)_{KFO} = \bigg(\frac{K^{*0}}{K}\bigg)_{CFO} \times e^{-\Delta t /\tau_{K^{*0}}}   \hspace{0.5cm}.
\end{equation} 
Here we have taken that the $(K^{*0}/{K})_{CFO}$ and $(K^{*0}/{K})_{KFO}$ are similar to the $K^{*0}/{K}$ ratios measured in elementary and heavy-ion collisions respectively. We have assumed that (i) there is no $K^{*0}$ regeneration taking place between the chemical and kinetic freeze out, and (ii) all $K^{*0}$ that decay before the kinetic freeze out are lost due to the re-scattering effect


In the right panel of Fig.~\ref{ fig-kstarbyk-ratio} we have shown the variation of $\Delta t$ as a function of $\sqrt{s_{NN}}$. Here we can see that measurements from RHIC seem to be smaller than that at the LHC, However, more statistics is needed in order to draw firm conclusions.

\section{Summary}

The production yield and azimuthal anisotropy measurements of (multi-)strange hadrons at STAR BES energies are reported. The rapid increase in $\Lambda/K^0_s$ ratio as a function of $p_T$ at 3 GeV could be due to the change in chemical potential at lower energies. The NCQ scaling of elliptic flow holds for the particles and corresponding anti-particles, which could be due to partonic collectivity. The suppression of $K^{*0}/K$ ratio suggests that there is a dominance of hadronic re-scattering in central heavy-ion collisions. Based on the $K^{*0}/K$ ratios, the extracted lifetime of the hadronic phase at RHIC seems to be smaller than that at the LHC.

\section{Acknowledgements}
Financial support from Department of Education, Government of India is gratefully acknowledged.

\bibliographystyle{unsrt}
\bibliography{template.bbl}


\end{document}